\documentclass[usenatbib]{mnras}

\usepackage{newtxtext,newtxmath}

\usepackage[T1]{fontenc}
\usepackage{ae,aecompl}

\usepackage{graphicx}	
\usepackage{amsmath}	
\usepackage{enumerate}
\usepackage[caption=false]{subfig}
\usepackage{calc}
\usepackage{comment}
\usepackage{threeparttable}
\usepackage{xcolor}
\usepackage{lipsum}

\newcommand{\pagefigurenv}[2]{\begin{figure*}
\centering
\includegraphics[width=\linewidth]{figures/#1}
\caption{#2}
\end{figure*}}

\title[Elemental Abundances in Low-Mass Galaxies]{Predictions for Complex Distributions of Stellar Elemental Abundances in Low-Mass Galaxies}

\author[Patel et al.]{Preet B. Patel,$^{1}$\thanks{E-mail: \href{mailto:pbpa@ucdavis.edu}{pbpa@ucdavis.edu}},
Sarah R. Loebman$^{1,2}$\thanks{Hubble Fellow},
Andrew Wetzel$^{1}$, \newauthor
Claude-Andr\'e Faucher-Gigu\`ere$^{3}$, 
Kareem El-Badry$^{4}$,
Jeremy Bailin${^5}$\\
$^{1}${Department of Physics \& Astronomy, University of California, Davis, 1 Shields Ave, Davis, CA 95616, USA}\\
$^{2}${Department of Physics, University of California, Merced, 5200 N. Lake Road, Merced, CA 95343, USA}\\
$^{3}${Department of Physics \& Astronomy and CIERA, Northwestern University, 2145 Sheridan Road, Evanston, IL 60208, USA}\\
$^{4}${Department of Astronomy and Theoretical Astrophysics Center, University of California Berkeley, Berkeley, CA 94720, USA}\\
$^{5}${Department of Physics and Astronomy, University of Alabama, Box 870324, Tuscaloosa, AL, 35487, USA}
}
\date{}

\pubyear{2021}

\begin{document}
\label{firstpage}
\pagerange{\pageref{firstpage}--\pageref{lastpage}}
\maketitle

\begin{abstract}
We investigate stellar elemental abundance patterns at $z=0$ in 8 low-mass ($M_{*}=10^{6}-10^{9}\,\text{M}_{\odot}$) galaxies in the Feedback in Realistic Environments (FIRE) cosmological simulations.
Using magnesium (Mg) as a representative $\alpha$-element, we explore stellar abundance patterns in [Mg/Fe] versus [Fe/H], which follow an overall monotonic trend that evolved slowly over time.
Additionally, we explore 3 notable secondary features in enrichment (in three different case-study galaxies) that arise from a galaxy merger or bursty star formation.
First, we observe a secondary track with a lower [Mg/Fe] than the main trend.
At $z = 0$, stars from this track are predominantly found within 2-6 kpc of the center; they were accreted in a 1:3 total-mass-ratio merger $\sim 0.4$ Gyr ago.
Second, we find a distinct elemental bi-modality that forms following a strong burst in star formation in a galaxy at $t_{\text{lookback}}$ $\sim 10$ Gyr.
This burst quenched star formation for $\sim 0.66$ Gyr, allowing Ia supernovae to enrich the system with iron before star formation resumed. 
Third, we examine stripes in enrichment that run roughly orthogonal to the dominant [Mg/Fe] versus [Fe/H] trend; these stripes correspond to short bursts of star formation during which core-collapse supernovae enrich the surrounding medium with Mg (and Fe) on short timescales.
If observed, these features would substantiate the utility of elemental abundances in revealing the assembly and star formation histories of dwarf galaxies. 
We explore the observability of these features for upcoming spectroscopic studies. 
Our results show that precise measurements of elemental abundance patterns can reveal critical events in the formation histories of low-mass galaxies. 
\end{abstract}



\section{Introduction} \label{sec:intro}
\lipsum[300]
In a hierarchical model of galaxy formation, low-mass galaxies are a fundamental building block of all of the higher-mass galaxies that we observe today \citep{Searle1978}.
As such, understanding how these low-mass structures formed and evolve provides indispensable insight into the early epochs of our own Galactic origin \citep[e.g.][]{Kirby2009p, Besla2010, El-Badry2018}.
One approach to understanding the star formation history of a galaxy is through studying the elemental abundances of its present day stellar constituents; in particular, characterizing the global trends in $\alpha$-to-iron ([$\alpha$/Fe]) versus iron-to-hydrogen ([Fe/H]) to probe both the epoch and environment of star formation. 
Here, `$\alpha$' refers to $\alpha$-element, or an element (O, Ne, Mg, Si, S, Ar, Ca and Ti) produced through the $\alpha$-capture process in stars \citep{Burbidge1958}.

Unfortunately, low-mass galaxies are typically low surface brightness and take up a small angular size on the sky, making it difficult to resolve individual stars within them.
Characterizing internal properties, like the radial distribution of stellar elemental abundances, and finding the global metal distribution is difficult in all but the most nearby galaxies \citep{Kirby2011}.
Progress on research in the nearby Universe is tied to developments in both observing tools and theory/simulations (discussed below).
In the past few decades, such developments have made tremendous progress in characterizing the star-formation histories and elemental abundances of individual stars for dwarf galaxies in the Local Group \citep[e.g.][]{Silk1987, Kirby2009, BlandHawthorn2010, Webster2016, Ji2018, Kirby2019, Sandford2020}.
Simultaneously, improvements in modeling methods and computational resources have resulted in a new generation of high-resolution cosmological simulations that capture realistic enrichment processes to the present day \citep[e.\,g.,][]{Brooks2018, Buck2021, Genina2019, Grand2018, Hopkins2018, Munshi2019}.
Thus, researchers can compare results from improved models with the newest observations in the local universe. 

For example, simulations now produce metallicity distributions consistent with observations of nearby low-mass galaxies in the Local Group. 
\citet{Escala2018} found that FIRE-2 reliably produced such metallicity distributions via the inclusion of sub-grid metal diffusion in galaxies \citep{Hopkins2018}.
This advance offers the potential to track the formation history of low-mass galaxies through both kinematics and elemental abundances.
In their analysis, \citet{Escala2018} consider the $\alpha$-versus-iron abundance patterns in four $\text{M}_{\text{halo}} = 10^{10} \, \text{M}_{\odot}$ systems.
The elemental abundance tracks follow a smooth, monotonic trend, indicating that these systems formed from mostly well mixed gas that self-enriched over time.
When compared to observations, these results are qualitatively consistent with dwarf galaxies in the Local Group, such as Leo A, Aquarius, and Sagittarius Dwarf-Irregular \citep{Kirby2017}.

Simulations continually push towards more accurate modelling of the evolution of elemental abundances in galaxies.
Earlier models, such as those in \citet{Pilyugin1993, Pilyugin1995}, found that the inclusion of galactic winds and self-enrichment is necessary to reproduce observed quantities, like Fe abundance distributions and the age-metallicity relation.
\citet{Romano2006} built on these principles to better model the star-formation histories of 2 dwarf galaxies (NGC1705 and NGC1569).

More recent works place greater emphasis on the trends in the $\alpha$-versus-Fe content in simulated galaxies.
Using the \textsc{Gasoline} simulations, \citet{Zolotov2010} analyzed the elemental abundances of simulated galaxies in the stellar mass range of $10^{9} - 10^{10} \, \text{M}_{\odot}$ (further details in their \S 2).
They compare stellar \textit{ex-situ} and \textit{in-situ} fractions in the [Mg/Fe] versus [Fe/H] in these systems (their Figure~3).
The \textit{ex-situ} stars correspond to a separate population of stars in elemental abundance space in 2 out of 3 cases.
They investigate the possibility of using these distinct populations in elemental abundance space to discern the role of accretion in galaxy formation, noting that \textit{ex-situ} stars are less $\alpha$-enhanced than \textit{in-situ} stars.

Using \textsc{Changa} (based on \textsc{Gasoline}), \citet{Governato2015} also investigated the $\alpha$-versus-Fe of simulated galaxies, identifying the `$\alpha$-knee' in their elemental abundance trends \citep{Tinsley1046}.
Most recently, \citet{Buck2021} generate galaxies (ranging from $M_{*} = 10^{6} - 10^{11} \, \text{M}_{\odot}$) with \textsc{Gasoline2} and vary several stellar evolution parameters to test their effects on present day galaxy properties.
They find a diversity in abundance patterns when looking at their simulated systems. 
While they do not investigate these trends in detail, they add that feedback and accretion may play major roles in forming this large diversity.

With the \textsc{Gear} code, \citet{Revaz2018} generate several low-mass galaxies in a $\Lambda$CDM universe and find that properties like star formation rates, mass-luminosity relationships, and metallicity gradients match observed dwarf galaxies well.
They find monotonic enrichment trends in the elemental abundance ratios ([Mg/Fe] versus [Fe/H]) of their simulated low-mass galaxies, much like the ones from \citet{Escala2018}.

In general, most simulation works to date have placed greater emphasis on the star-formation history, radial distributions of abundances, the age-metallicity relation, metallicity distribution, and kinematics in low-mass galaxies.
Authors rarely investigate trends in [$\alpha$/Fe] versus [Fe/H] of lower-mass galaxies.
With \textsc{Gasoline}, \citet{Zolotov2010} show that the elemental abundances of galaxies with $M_{\text{total}} \approx 10^{11} \, \text{M}_{\odot}$ contain distinct groupings for \textit{in-situ} and \textit{ex-situ} stars.
We aim to expand on similar results in FIRE-2 to see what such complexity can tell us about a galaxy's accretion and formation history, even at the low-mass end.

Observationally, many works have explored elemental abundances in low-mass galaxies.
The Large Magellanic Cloud (LMC) is a nearby galaxy that recently underwent a burst of star formation due to its interactions with the Small Magellanic Cloud (SMC).
Measurements of the $\alpha$-abundances of stars within the LMC are growing with time \citep[for example][]{Pompeia2008,VanderSwaelmen2013,Sakari2017, Nidever2020}, providing an increasing sample size with greater precision.
\citet{Kirby_2018} sample nearly 1,000 stars in Local Group dwarf galaxies as well.
Furthermore, upcoming spectroscopic data from James Webb Space Telescope (JWST) and the Extremely Large Telescopes (ELTs) will continue to enhance the resolution of elemental abundances in nearby dwarf galaxies \citep{deZeeuw2014, Gardner2006, Sandford2020}.
The amount of high-resolution spectroscopic data continues to improve, and we are presented with a growing opportunity to confidently compare simulations with observations. 
Therefore, understanding the $\alpha$-versus-iron content of simulated low-mass galaxies will provide insight into their galactic origins.

In this work, we analyze the [$\alpha$/Fe] versus [Fe/H] space of several low-mass galaxies simulated with the FIRE model, and we identify notable features.
In \S\ref{sec:sim}, we introduce the FIRE-2 simulations that we use.
In \S\ref{sec:res}, we examine the overall trends in elemental abundances.
We then identify the source of striations in the elemental abundance space in one galaxy with $M_{*} = 10^{8} \, \text{M}_{\odot}$, and the source of a bimodality in the [Mg/Fe] versus [Fe/H] of another galaxy with $M_{*} = 10^{7} \, \text{M}_{\odot}$.
Furthermore, we identify the origins of a secondary track observed in one of the galaxies with $M_{*} = 10^{6}M_{\odot}$.
In \S\ref{sec:Observability}, we explore the implications of our results the feasibility of observing them.
In \S\ref{sec:conc}, we summarize and discuss our findings.

\begin{table*}
\centering
\renewcommand{\thempfootnote}{\arabic{mpfootnote}}
\caption{
Properties (at $z = 0$) of the simulated galaxies in this work.
$M_{\text{200m}}$ refers to the total mass of gas, stars, and dark matter enclosed within the galaxy's virial radius ($R_{\text{200m}}$).
$\text{Res}$ refers to the mass resolution of each star/gas particle in the simulation in solar masses ($\text{M}_\odot$).
We define \textit{ex-situ} to be stars that formed $> 10$ kpc from the center of the galaxy. Superscript indicates the work that introduced each galaxy at this resolution.}
\begin{tabular}{r|c|c|c|c|c|c|c}
\hline
simulation & $M_{*}$ [$\text{M}_\odot$] & $M_{\rm gas}$ [$\text{M}_\odot$] & $M_{\text{200m}}$ [$\text{M}_\odot$] & $\text{R}_\text{200m}$ [kpc] & $\langle[ \text{Fe}/\text{H}]\rangle$ & Res [$\text{M}_\odot$] & ex-situ \% \\ \hline

m11d$^2$ & 4.6 $\times 10^{9} $&  1.1 $\times 10^{11} $ & 7.4 $\times 10^{11} $& 52 & -0.92 & 7100 & 15.5\% \\
m11e$^2$ & 1.6 $\times 10^{9} $&  7.8 $\times 10^{10} $ & 5.2 $\times 10^{11} $& 37 & -1.1 & 7100 & 37.1\% \\
m11i$^2$ & 1.0 $\times 10^{9} $&  3.2 $\times 10^{10} $ & 2.1 $\times 10^{11} $& 31 & -0.94 & 7100 & 6.0\% \\
m11q$^1$ & 4.7 $\times 10^{8} $&  6.6 $\times 10^{9} $  & 1.6 $\times 10^{11} $& 24 & -1.6 & 880 & 12.0\% \\ 
m11h$^2$ & 1.9 $\times 10^{8} $&  3.3 $\times 10^{9} $  & 7.3 $\times 10^{11} $& 16 & -1.6 & 880 & 19.5\% \\
m11b$^3$ & 4.7 $\times 10^{7} $&  2.7 $\times 10^{10} $ & 1.6 $\times 10^{11} $& 12 & -1.9 & 2100 & 24.2\% \\ 
m10q$^1$ & 2.5 $\times 10^{6} $&  4.1 $\times 10^{9} $  & 2.5 $\times 10^{10} $& $<$ 10 & -2.3 & 250 & 3.4\% \\
m10v$^1$ & 1.9 $\times 10^{6} $&  9.6 $\times 10^{9} $  & 5.8 $\times 10^{10} $& $<$ 10 & -1.5 & 250 & 0.0\% \\

\hline
\label{table:1}
\end{tabular}
\begin{tablenotes}
            \item[1] $^1$ \citet{Hopkins2018}, $^2$ \citet{El-Badry2018}, $^3$ \citet{Chan2018}
        \end{tablenotes}
\end{table*}

\section{FIRE-2 Simulations} \label{sec:sim}
Our simulations use the code \textsc{Gizmo} \citep{Hopkins2015} and the FIRE-2 physics model \citep{Hopkins2018}.
\textsc{Gizmo} uses a Tree-PM gravity solver from GADGET-3 \citep{Springel2005} and models hydrodynamics via the Lagrangian mesh-free finite-mass (MFM) method.
FIRE-2 includes radiative heating and cooling from atomic, molecular, and metal-line channels, including 11 elements (H, He, C, N, O, Ne, Mg, Si, S, Ca, Fe), across $10 - 10^{10}$ K.
This includes free-free, photoionization/recombination, Compton, photoelectric and dust collisional, cosmic-ray, metal-line, molecular, and fine-structure processes, as well as ionization and heating from a redshift-dependent, spatially uniform ultraviolet background \citep{Faucher2009}.
Appendix B of \citet{Hopkins2018} contains additional information on the cooling tables.
Star formation occurs in self-gravitating, Jeans-unstable, molecular gas that reaches a critical density ($n_{\rm SF} > 1000 ~ \text{\text{cm}}^{-3}$).
Individual star particles represent a single stellar population, assuming a \citet{Kroupa2001} stellar initial mass function.

Implemented in FIRE-2 are several stellar feedback channels, including (1) radiation pressure, (2) energy, momentum, mass and metal injection from supernovae and stellar mass loss (core-collapse rates from \citet{Leitherer2011}, Ia rates from \citet{Mannucci2006}, and (3) photoionization and photoelectric heating, drawing appropriate values for mass, momentum, and thermal energy injection from STARBURST99 v7.0 \citep{Leitherer2011}.
Our nucleosynthetic yields follow \citet{Iwamoto1999} for Type Ia supernovae (SNe Ia) and follow \citet{Nomoto2006} for core-collapse supernovae (CCSN). 
Stellar wind yields are sourced from a compilation of \citet{Groenewegen1997}, \citet{Marigo2001}, and \citet{Izzard2004}.
See \citet{Hopkins2018} for more comprehensive details on the FIRE-2 model.

We examine a subset of the cosmological zoom-in simulations from the FIRE-2 project \citep{Hopkins2018}; this suite has a resolution limit of 1 pc (adaptive) spatial in gas and $250M_{\odot} \ - \ 7100M_{\odot}$ in mass.
We choose this subset, because it corresponds to the mass range of nearby dwarf galaxies with well sampled observed stellar populations that show extended star formation histories (e.g. LMC, SMC, Leo P).
Thus, they are more likely to exhibit complex abundance distributions. 
We scale the values for abundances in stars to the (proto)solar abundances in \citet{Asplund2009}.

These \text{FIRE-2} simulations also explicitly model the sub-grid diffusion of metals in gas via turbulence, as \citet{Hopkins2018} describes in detail; \citet{Escala2018} showed how the inclusion of this model significantly improved the agreement of elemental abundance distributions in FIRE-2 low-mass galaxies as compared with observations.
The ability to resolve diffusion at this scale gives rise to the formation of singular, cohesive tracks in the $\alpha$-abundance space of simulated galaxies.
While \citet{El-Badry2016} detailed the kinematics of some of these galaxies with halo masses of $10^{11} M_{\odot}$, we consider the elemental abundances.

We identify (sub)halos using the ROCKSTAR 6D halo finder \citep{Behroozi2012a}.
This includes all (sub)halos that contain a bound mass fraction greater than 0.4 and at least 30 dark-matter particles within a radius of 200 times the mean matter density ($R_{\text{200m}}$).
For more details, see \S2.2 in \citet{Samuel2020}.

We focus on 8 galaxies with total halo mass $M_{\rm 200m} = 10^{10} - 10^{11} M_{\odot}$ and with stellar masses of $10^{6} - 10^{9} M_{\odot}$.
Table~\ref{table:1} highlights the halo mass ($M_{\rm total}$), gas mass ($M_{\rm gas}$), stellar mass ($M_{*}$), average [Fe/H], mass resolution, ex-situ fraction (selecting `ex-situ' as $R_{\text{form}} \geq 10$ kpc), and the paper that introduced each simulated galaxy at the given resolution.
We evaluate the metallicity ([Fe/H]) versus the $\alpha$-abundances (represented by [Mg/Fe]) of stars at present day, with additional considerations for the age of stars, their radius of formation, and the merger history of the galaxy.

\pagefigurenv{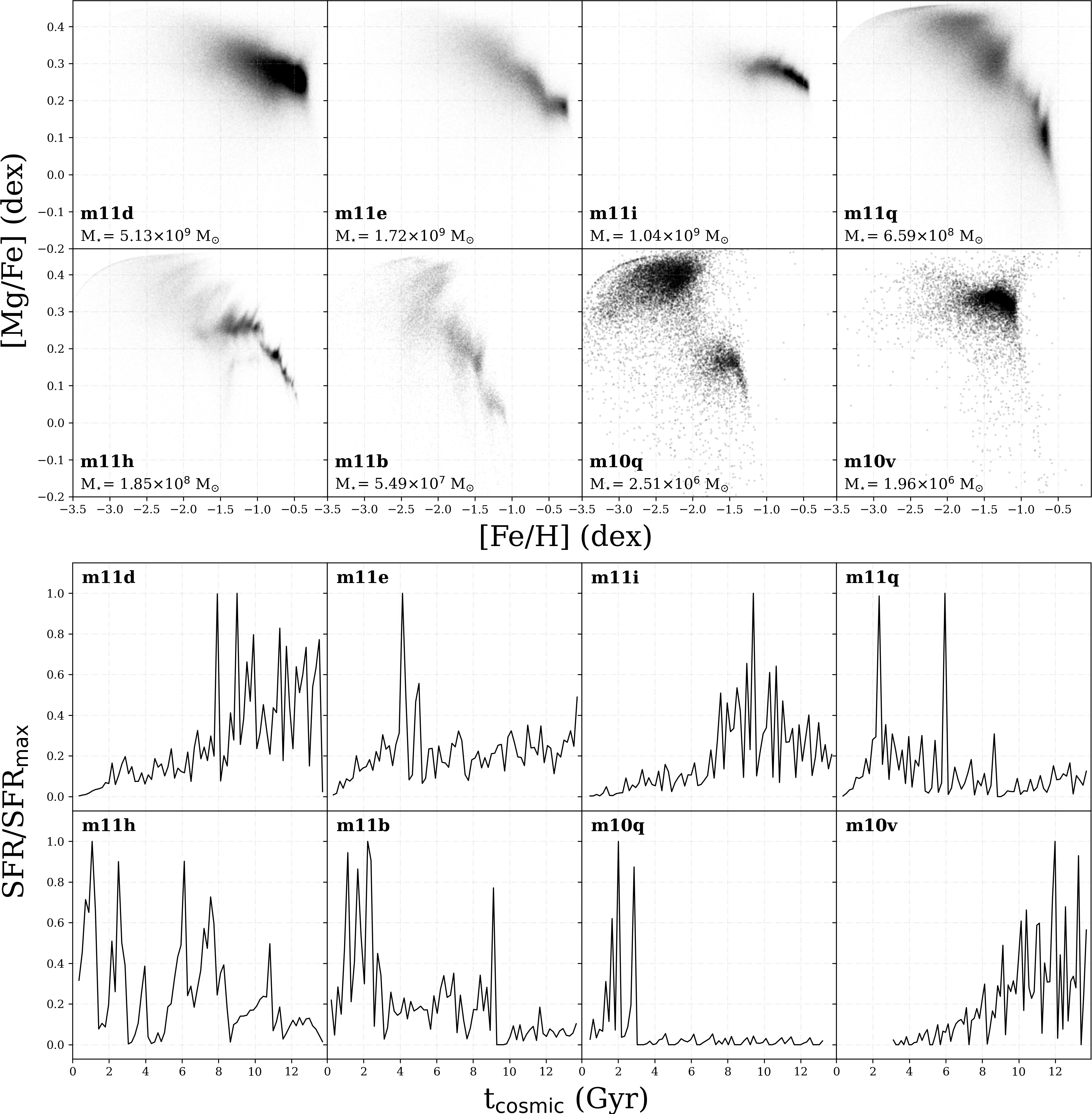}{
\label{fig:global_scatter}
\textit{Top}: [Mg/Fe] versus~[Fe/H] at $z = 0$ for star particles within 15 kpc of the center of our 8 simulated galaxies.
From top left to bottom right we present the [Mg/Fe] versus [Fe/H] for all galaxies in Table~\ref{table:1}, ordered by stellar mass. 
We present star particles as points to distinguish clearly the key features that we explore (which are obscured in a 2-D histogram/heatmap).
Stars in all 8 systems follow the expected overall trend: high [Mg/Fe] at low [Fe/H] to low [Mg/Fe] at higher [Fe/H].
\textit{Bottom}: Star-formation rate (SFR), normalized to the maximum SFR across the history of each galaxy.
The peaks in SFR signify periods of rapid star formation or `bursts'. 
Each galaxy differs in the number of peaks, as well as their strength and frequency. 
Several galaxies (\textbf{m11d, m11q, m11h, m11b, m10v}) have star formation histories with several peaks throughout time.
The other systems contain similar peaks as well, but have fewer in total (\textbf{m11e, m11i, m10q}).
Overall, we find evidence of bursty star formation and accretion in \textbf{m11b} and \textbf{m11q} (via stripes in elemental abundances), \textbf{m10q} (via a break in elemental abundances), and \textbf{m11h} (via a secondary track in elemental abundances), which we explore in \S\ref{m11h} - \S\ref{m11b}.}

\section{Results}
\label{sec:res}

\subsection{Overall Trends in Abundances}
\label{m11i}

We begin by considering the overall patterns in elemental abundances for our simulated galaxies.
FIRE-2 simulations track a variety of $\alpha$-elements, including magnesium (Mg), calcium (Ca), and silicon (Si).
We examine Mg as a representative $\alpha$-element, in part because, within the FIRE-2 model, core-collapse supernovae most cleanly dominate its total yield.
We find qualitatively similar trends analyzing [O/Fe], [Si/Fe], and [Ca/Fe] as well.

First, Figure~\ref{fig:global_scatter} (top) shows the [Mg/Fe] versus [Fe/H] for stars at $z = 0$ in each galaxy.
Each point represents a single star particle.
We display individual points (instead of using a heatmap/2-D histogram) to make the structures/features in this abundance space as qualitatively clear as possible.

Otherwise, we refer to the relatively smooth, monotonic trend that the rest of the stars generally fall along as the `elemental abundance trend', which corresponds to how the system steadily self-enriches over time \citep{Searle1972, Ji2018}. 
Observationally, dwarf galaxies like Fornax and Leo I show such a trend in their distributions of [$\alpha$/Fe] versus [Fe/H] \citep{DeBoer2012, Vargas2014}.

Throughout, we limit our analysis to [Fe/H] $> -3.5$; below this, the metallicity floor of our simulations leads to an artificial effect on the abundances (faintly seen in the top left corners of each panel).
Specifically, the first stars in these simulations formed at an imposed floor of [Fe/H] $=$ [Mg/H] $\approx -3.8$, so with [Mg/Fe] $= 0$, and as star formation proceeded, this rose to a ceiling of [Mg/Fe] $\approx 0.45$, which reflects the ratio of Mg production relative to Fe during CCSN.

Figure~\ref{fig:global_scatter} (bottom) shows the star formation rate (SFR) of each galaxy relative to its maximum.
Galaxies like \textbf{m11d}, \textbf{m11i}, and \textbf{m10v} had SFRs that peaked closer to present day than the other galaxies.
\textbf{m11h}, \textbf{m11b}, \textbf{m11e}, and \textbf{m10q} all show burstier star formation throughout their lifetimes, with higher SFRs in the past.
\textbf{m10q} is an extreme case, where star formation peaked at 3 Gyr.
This peak corresponds to a break within its overall abundance distribution.
In fact, peaks in the SFR directly correspond to features in the abundances of several galaxies in which the SFR was generally higher in the past (\textbf{m11q}, \textbf{m11h}, \textbf{m11b}, \textbf{m10q}).
That is not to say that bursty star formation necessarily leads to featuring in the abundances.
Galaxies with SFRs that skew much later tend to not show these features (\textbf{m11d}, \textbf{m11e}, \textbf{m11i}, \textbf{m10v}). 
We find that in \textbf{m11b}, \textbf{m11q}, and \textbf{m10q}, the bursty SFRs relate to striations or breaks in the abundance plane, as we explore below.

In our simulated galaxies, at [Fe/H] of $-3.5$ the $1-\sigma$ scatter of [Mg/Fe] around the mean is $\sim 0.1 - 0.2$ dex.
This value steadily drops to $\sim 0.05$ dex at [Fe/H] of -0.5.
For all galaxies, the average $1\sigma$ scatter in [Mg/Fe] around the mean is $\sim 0.1$ dex and becomes smaller with time.
This is what we refer to when an elemental enrichment trend follows a `tight' distribution.
\citet{Escala2018} investigate the scatter of abundance distributions in FIRE-2, and find that low-mass galaxies are generally well mixed at all times. This results in low scatter in [$\alpha$/Fe] (0.05-0.1 dex).
Furthermore, \citet{Escala2018} show that this small scatter is consistent with observations, once one incorporates observational uncertainties.
\citet{Muley2020} study the effects of turbulent diffusion of metals and varying nucleosythentic yields on the intrinsic scatter in [Mg/Fe] and [Fe/H], and find that this diffusion is necessary to reproduce the observed scatter.


Of the 8 galaxies we examine, 4 (\textbf{m11d}, \textbf{m11i}, \textbf{m11q}, \textbf{m10v}) are relatively smooth, monotonic, and feature-less; they do not contain the features in their abundances that we investigate.
We include these systems to provide the reader with a sense of the variety in abundance distributions within our available simulations. 
Generally, the distribution in [Mg/Fe] versus [Fe/H] for these galaxies, such as \textbf{m11i}, is self-similar at all radii (not shown), varying $< 0.1$ dex across all radii.
Such results are consistent with a bursty star formation model, indicating that global potential fluctuations drove stars to dispersion-dominated orbits with time, as \citet{El-Badry2016} showed in FIRE simulations.
While the qualitative shape of the abundance tracks are relatively self-similar at all radii, the normalization of the metal abundance tracks in the [Mg/Fe] versus [Fe/H] plane shifts towards a higher metallicity in higher-mass systems, as expected.

Some of these galaxies that appear smooth in [Mg/Fe] versus [Fe/H] in fact contain features, but they are not visible in the overall distribution.
For example, \textbf{m11e} contains a significant accreted fraction of stars (37\%, see Table~\ref{table:1}), but we do not observe a distinct secondary feature containing them.
We identify a separate track in \textbf{m11e} using the $r_{\text{form}}$ of stars (not shown), but it is indistinguishable from the main trend.
As a result, we do not investigate such `features' in detail.

The remaining 4 of 8 (half) galaxies in our sample contain distinct features in their enrichment trends.
Varying the radial bin in which we select stars for these galaxies can affect the visibility of abundance features (see \S\ref{m11h}); however, they are still visible across most radii at present day.
When investigating the [$\alpha$/Fe] versus [Fe/H], we see multiple tracks in \textbf{m11h} \& \textbf{m10q}, alongside visible striations in \textbf{m11b}, and \textbf{m11h}.

Figure~\ref{fig:global_scatter} shows a subtle trend with mass, that our lower-mass galaxies show somewhat stronger visible features in their abundance patterns.
One possible concern, as Table~\ref{table:1} shows, is that our simulations have varying resolution, from $7100$ to $250 \text{M}_\odot$, with lower resolution for higher-mass galaxies.
So, we examined the potential resolution dependence of these results by analyzing versions of \textbf{m11q} and \textbf{m11h} at $8 \times$ lower resolution.
In general, we find the same features in the lower-resolution simulations (not shown), though with significant stochasticity (as driven by stochasticity in the formation history).
For example, \textbf{m11q} shows stripes/striations in its abundance patterns, and at lower resolution ($7100 \text{M}_\odot$) it contains even more of these striations.
Conversely, we find the opposite trend for \textbf{m11h}: the lower-resolution simulation shows weaker features in the secondary tracks that we explore in \S~\ref{m11h}.
Thus, we conclude that resolution does not play a significant systematic role in presence of these features.

We identify 3 key physical processes that lead to these features: satellite mergers, bursty star formation with quick enrichment, and a large individual burst that pauses star formation for extended periods of time, permitting Fe enrichment from Ia supernovae.
We explore these features further in \S\ref{m11h} - \S\ref{m11b} but briefly summarize them as follows.
\textbf{m11h} shows a secondary track in the [Mg/Fe] versus [Fe/H] from a galaxy merger in its recent history.
\textbf{m10q} shows a break in [Mg/Fe] versus [Fe/H] from a strong burst in star formation.
Finally, \textbf{m11b} shows stripes in its [Mg/Fe] versus [Fe/H] from rapid Mg enrichment during bursts in star formation; the process is recurring and results in multiple instances of this feature.

\pagefigurenv{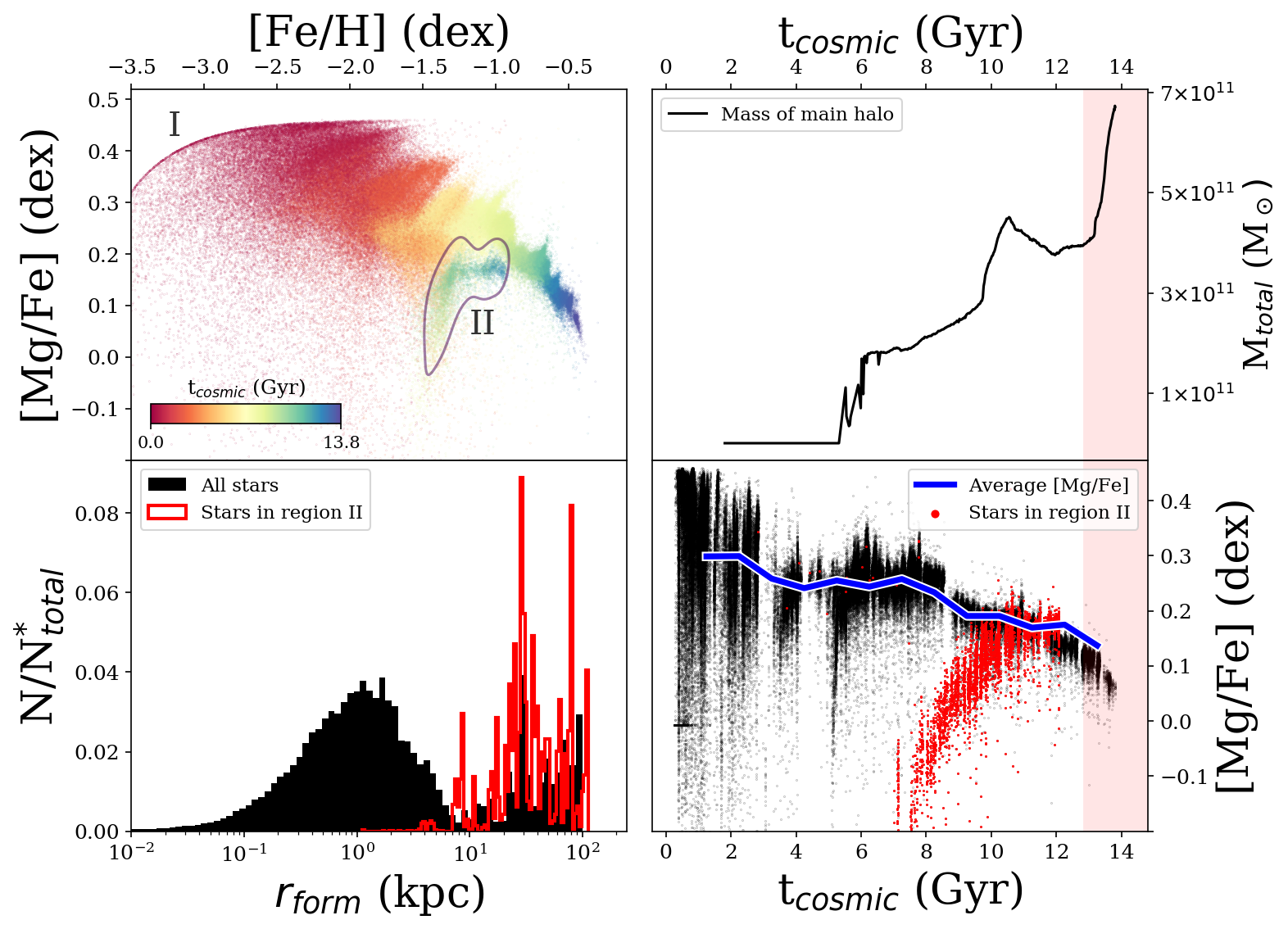}{
\label{fig:m11h_master}
Key aspects of the formation history of \textbf{m11h}, a case study for the imprint of a merger on a distinct enrichment track.
\textit{Top left}: [Mg/Fe] versus [Fe/H], with color mapped to the cosmic time at which stars formed.
Encircled is a secondary track, chromatically indicated by its relatively young stars.
\textit{Top right}: the total mass of the host halo as a function of time.
The pink region conveys the epoch of a 1:3 total mass merger.
\textit{Bottom left}: The radius at which stars formed (distance from progenitor halo at the time of formation). 
Both sets are normalized to their respective total counts.
\textit{Bottom right}: After this highlighted period, the average [Mg/Fe] drops significantly.
Furthermore, the stars in the secondary track all have high iron abundance and formed prior to the merging event.
Thus, the stars that make up the secondary track formed recently and were accreted in a merger.
}

\pagefigurenv{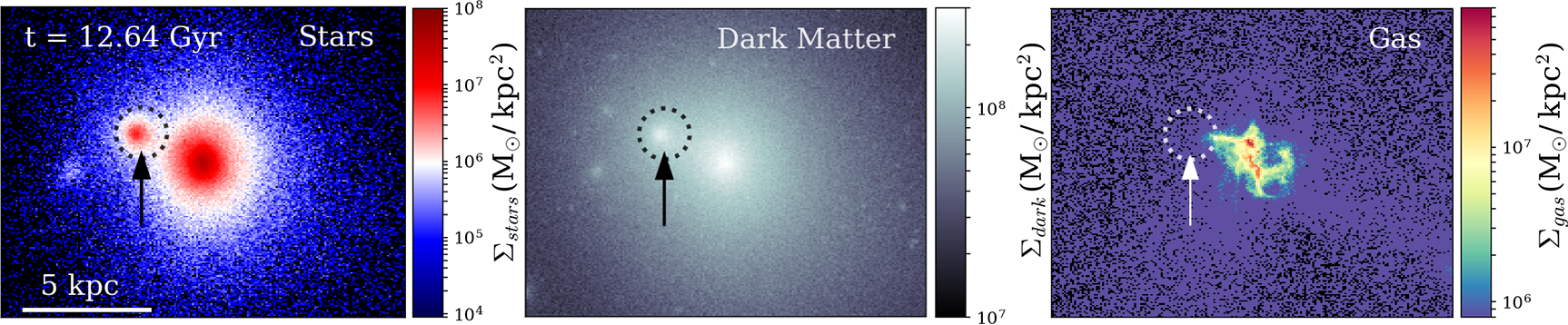}{
\label{fig:m11h_3column}
Images of stars, dark matter, and gas in m11h at $t_{\text{cosmic}} = 12.64$ Gyr ($z = 0.09$), when another galaxy, largely devoid ($\leq 10^6 M_{\odot}/\text{kpc}^2$) of gas, fell into the halo.
At this time, this satellite has a total mass of roughly $1.7 \times 10^{11} M_{\odot}$ and a stellar mass of $6.71 \times 10^{6} M_{\odot}$.
All of the stars that make up the secondary track in \textbf{m11h} came from this accreted galaxy.
}

\subsection{Secondary Track in m11h from Galaxy Accretion}
\label{m11h}

We first examine how a galaxy merger can generate a secondary track in [Mg/Fe] versus [Fe/H], using \textbf{m11h} as a case study.
Figure~\ref{fig:global_scatter} (panel 8) shows a distinct feature in the present-day abundances of \textbf{m11h}.
Figure~\ref{fig:m11h_master} (top left) shows [Mg/Fe] versus [Fe/H] for stars; we indicate their cosmic time of formation via color: red corresponds to the oldest stars and blue corresponds to the youngest.
The ages of stars in region II are offset from the ages of stars in a similar bin of [Fe/H].

To understand the origin of these stars, Figure~\ref{fig:m11h_master} (bottom left) shows the radius of formation for all stars in \textbf{m11h} (black histogram) and stars from region II (red histogram).
We normalize all values to the total population of consideration.
Region II stars formed $>$10 kpc from the host, and account for roughly 3\% of the stars in the galaxy.
Given their distinct abundances and distances of formation, region II stars formed in a different environment than the majority of stars in the main galaxy. 
Figure~\ref{fig:m11h_master} (top right) shows the total halo mass as a function of time (black line).
This galaxy experienced 2 periods ($t_{\text{cosmic}} \approx 10 \ \text{Gyr}$ and $\sim 12 \ \text{Gyr}$) of rapid mass gain in its history.
We do not consider the first period, because it predates the formation of most stars in region II.
We highlight the second period in red between 12 Gyr and present day, when the mass of the host halo increased by $\sim 1.7 \times 10^{11} M_{\odot}$ to a total mass of $\sim 5.1 \times 10^{11} \text{M}_{\odot}$ at present day. 
Figure~\ref{fig:m11h_master} (bottom right) shows [Mg/Fe] as a function of cosmic time for all stars in the galaxy.
Region II stars (red) have a different [Mg/Fe] from stars of a similar age.
These stars also formed prior to the period of rapid accretion near the end of \textbf{m11h}'s history.

In short, stars from region II formed together, far ($\sim 30$ kpc) from the main galaxy, and at a distinctly different [Mg/Fe] from stars of a similar age and metallicity.
The formation of these stars predated an epoch of rapid mass growth in the galaxy.


Using the halo catalogs, we confirm that the secondary track encircled in Figure~\ref{fig:m11h_master} (top left panel) corresponds to a galaxy with $M_{\text{total}} = 1.7 \times 10^{11} M_{\odot}$ accreted $\sim 400$ Myr ago.
Figure~\ref{fig:m11h_3column} depicts an early passage of this galaxy; it contained stars (left, $M_{\text{star}} = 6.7 \times 10^6 M_{\odot}$), a dark-matter halo (middle, $M_{\text{dark}} \approx 1.7 \times 10^{11} M_{\odot}$), and was largely devoid of gas (right, $M_{\text{gas}} \approx 1 \times 10^6 M_{\odot}$) before the merger.
The stellar mass ratio between the merger and host is roughly 1:40, and the total mass ratio is roughly 1:3.
$\sim$23\% of stars in \textbf{m11h} formed ex-situ ($r_{\text{form}} > ~10$ kpc), and region II stars make up 13\% of all ex-situ stars.
The relatively low gas mass supports the notion that region II stars from Figure~\ref{fig:m11h_master} formed prior to accretion, and that the merger deposited them throughout the main galaxy.
The satellite hosting these ex-situ stars never interacted with the main halo - it fell in from a large distance (not shown). 
We find remnants of the fossil system throughout the galaxy as a feature in the overall [$\alpha$/Fe] versus [Fe/H].

Figure~\ref{fig:m11h_radial} shows the radial dependence of this secondary track.
Each panel displays stars within a 2 kpc radial bin, with 4 bins spanning 0 kpc to 8 kpc.
We choose this range given that \textbf{m11h}'s stellar $R_{\text{90}} = 8.4$ kpc.
We outline the secondary track in red.
While we find stars from the region in all four panels, this track is most distinct from the main trend between 2 kpc and 4 kpc, and is qualitatively least visible between 6 kpc and 8 kpc.
Thus, radial variation does not significantly alter the visibility of the secondary track and main trend within 6 kpc. 
In the right column, the top panel shows the [Fe/H] of \textbf{m11h} for stars in the same selections.
The highest-metallicity stars are at smaller radii ($< 4$ kpc), while the most metal-poor stars are found in larger radial bins (4-8 kpc).
\citet{Graus2019} and \citet{Mercado2020} investigated the radial dependencies of FIRE-2 dwarf galaxies and characterized their radial age variations and mean metallicity gradients.
Stars in central regions of these galaxies tend to be younger and metal-rich compared to stars at larger radii.
The mean abundances for our galaxies generally agree with these analyses (example in \S\ref{m11h}).
However, such radial trends are not the focus of our analysis, because we investigate the variations in the \textit{overall} abundance distributions.
The bottom panel shows the [Mg/Fe] for stars in the same radial bins. 
Here, we find the opposite relation - the lowest-[Mg/Fe] stars occupy the smallest radial bins, and we find more enriched populations at larger radii.
Figure~\ref{fig:m11h_radial} demonstrates how the imprint of accretion varies with radius, though it is generally present at all radii.
In fact, all features that we examine in this work persist at all radii within a galaxy, though with varying strength.

In summary, we observe a secondary track in the present-day elemental abundances of \textbf{m11h},
arising as the the fossil remnant of a single merger event.
This feature shows little variation with radius, though it is more pronounced in the inner galaxy. 
Thus, this secondary trend in \textbf{m11h}'s elemental abundances indicate a merger in its formation history.


\pagefigurenv{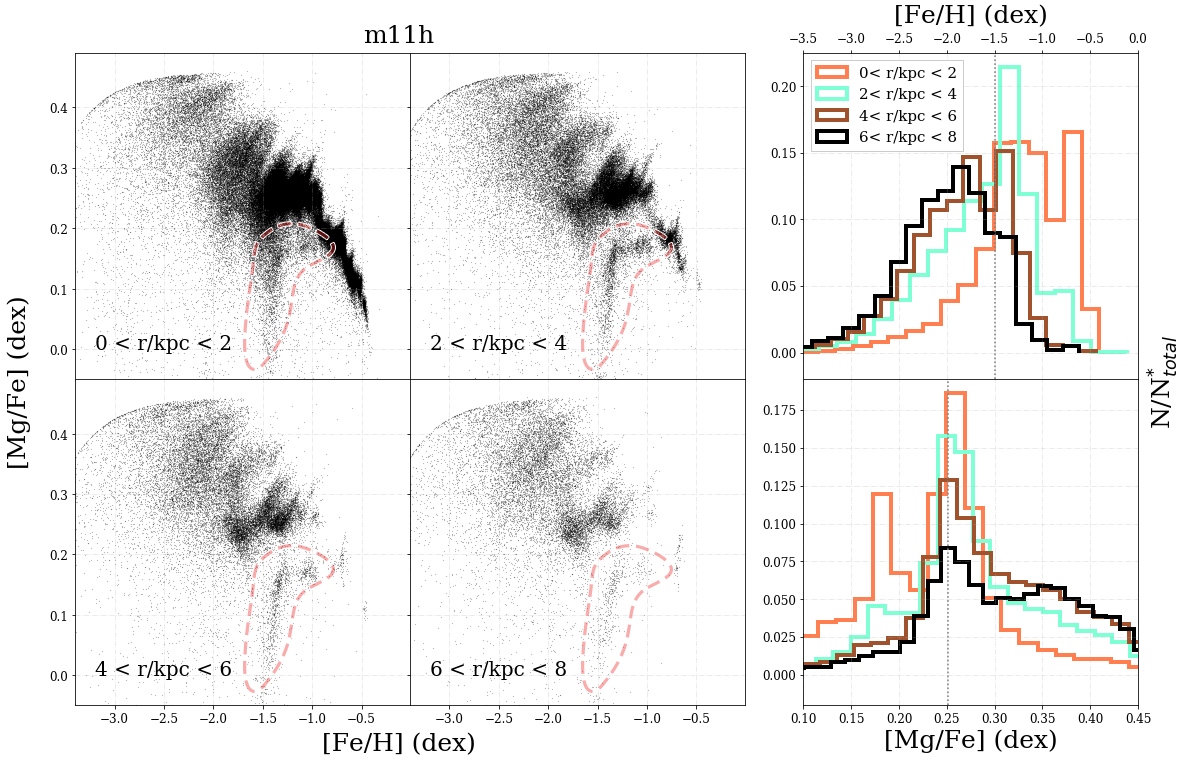}{\label{fig:m11h_radial}
\textit{Left}: [Mg/Fe] versus [Fe/H] at $z = 0$ for star particles in \textbf{m11h} in 4 radial bins: 
$0 - 2$ kpc (top left), 
$2 - 4$ kpc (top right), 
$4 - 6$ kpc (bottom left), 
$6 - 8$ kpc (bottom right).
The secondary track is faintly outlined in red. 
In \textbf{m11h}, $R_{\rm 90} \approx 8.4$ kpc. 
\textit{Right}: distributions for all 4 radial bins in [Fe/H] (top) and [Mg/Fe] (bottom).
The the gray dashed line shows the average [Fe/H] (-1.5) and the average [Mg/Fe] (0.25) in both panels.
From the central radial bins, we see that stars within the inner regions of the galaxy have higher metallicities than stars at larger radii.
\textbf{m11h} contains a secondary feature in its [Mg/Fe] versus [Fe/H] that could signify events in the galaxy's history.
This feature is qualitatively the most distinct $2 - 4$ kpc from the center, but we observe stars from the region across all 4 radial bins.
}

\subsection{Secondary Track in m10q from Bursty Star Formation}
\label{m10q}

Figure~\ref{fig:m10q_master} (top left) shows [Mg/Fe] versus [Fe/H] for \textbf{m10q}.
As in Figure~\ref{fig:m11h_master}, each point represents a star particle and its color indicates its cosmic time of formation.
We find a $\sim 0.3$ dex drop in the average [Mg/Fe] of stars at [Fe/H] $\approx -2.0$, unique to \textbf{m10q} in our suite (Figure~\ref{fig:global_scatter}).
We refer to this offset as the `break' in the elemental abundance trend.
We divide the stars into 2 regions: stars that formed prior (top left) are in region I, and the stars that formed after the break (bottom right) are put in region II.
The youngest stars in region I formed at $t_{\text{cosmic}} \approx 2.9$ Gyr.
The stars in region II formed after stars that formed in region I, at a higher [Fe/H], and at a lower [Mg/Fe].
These stars comprise $\sim$20\% of the total number of stars in \textbf{m10q} and constitute the vast majority (90\%) of stars that formed after the break.

\pagefigurenv{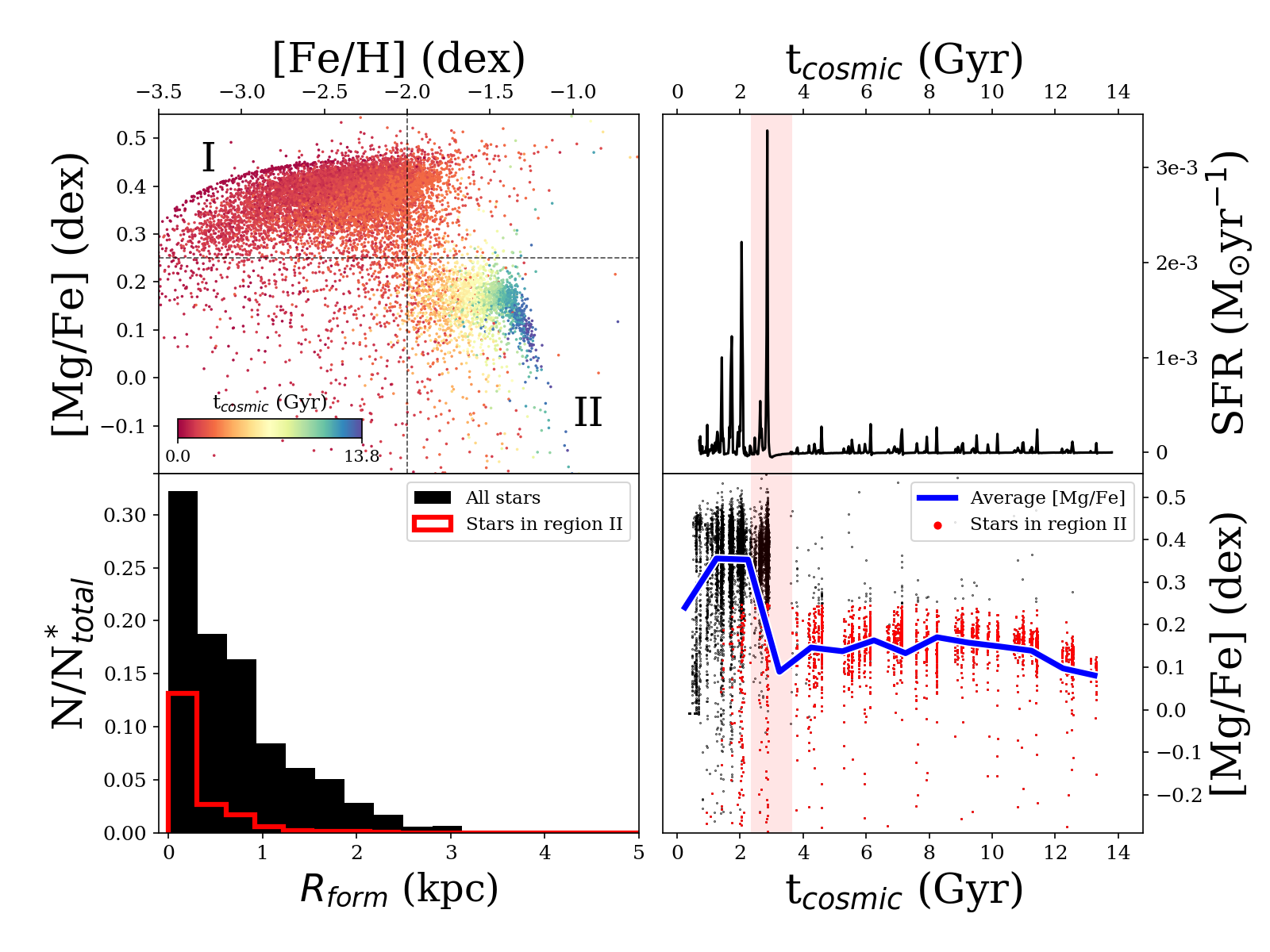}{\label{fig:m10q_master}
Key aspects of the formation history of \textbf{m10q}, a case study of how bursty star formation imprints distinct enrichment tracks.
\textit{Top left}: [Mg/Fe] versus [Fe/H], color-coded by time of formation (red/blue for early/late forming stars, respectively).
Dotted lines guide the eye; regions containing the majority of star particles labeled regions I and II.
Note the clear break and downward offset in [Mg/Fe] for star particles forming after 4 Gyr (in region II).
\textit{Bottom left}: formation distance relative to the center of {\tt m10q}; all star particles from region I and II formed within a few kpc of the center of the galaxy.
\textit{Top right}: star-formation rate of {\tt m10q} as a function of cosmic time.
Pink region is used to highlight a period after which the average [Mg/Fe] dropped significantly.
\textit{Bottom right}: [Mg/Fe] versus time of formation for stars in {\tt m10q}.
In light red, we highlight a period when the average [Mg/Fe] (blue line) dropped significantly ($\sim 0.25$ dex).
This `break' in [Mg/Fe] versus [Fe/H] correlates with a burst in star formation at $\sim 3$ Gyr.
}

Figure~\ref{fig:m10q_master} (bottom left) contains a histogram of the radius of formation for all the stars (black) in \textbf{m10q}, as well as a histogram for the stars in region II (red).
We normalize each histogram to the total number of stars in each respective population.
Unlike \textbf{m11h}, these stars formed in-situ within the main galaxy (were not accreted).
Additionally, we found no indicators of rapid mass gain in \textbf{m10q}, and no mergers whose stars corresponded with the stars in region II; we conclude that these stars formed \textit{in-situ} ($R_{\text{form}} < 10$ kpc).

We evaluate the instantaneous star-formation rate (black) as a function of cosmic time (Figure~\ref{fig:m10q_master}, top right).
The pink highlighted region at $t_{\text{cosmic}} \approx 3 Gyr$ indicates a strong burst of star formation across a $\sim 0.2$ Gyr window of time.
During this period, the stellar mass increased by $M_{*} = 5.1 \times 10^{5} \, \text{M}_{\odot}$, and no stars formed for $\sim 0.66$ Gyr after.
This burst accounts for nearly 20\% of star particles in the galaxy at present day, with a total stellar mass of $3 \times 10^9 \, \text{M}_{\odot}$.
Figure~\ref{fig:m10q_master} (bottom right) shows the [Mg/Fe] of all stars (black) as a function of cosmic time.
We show region II stars in red, and find that they formed steadily after the pause in star formation until present day.
After this, new stars form at a lower [Mg/Fe] and a higher [Fe/H]; we find that the burst in star formation resulted in feedback that displaced gas (and stars), stifling star formation.
Why then, does the overall [Mg/Fe] drop and the [Fe/H] increase after star formation resumes, despite no new gas being added to the system?

\pagefigurenv{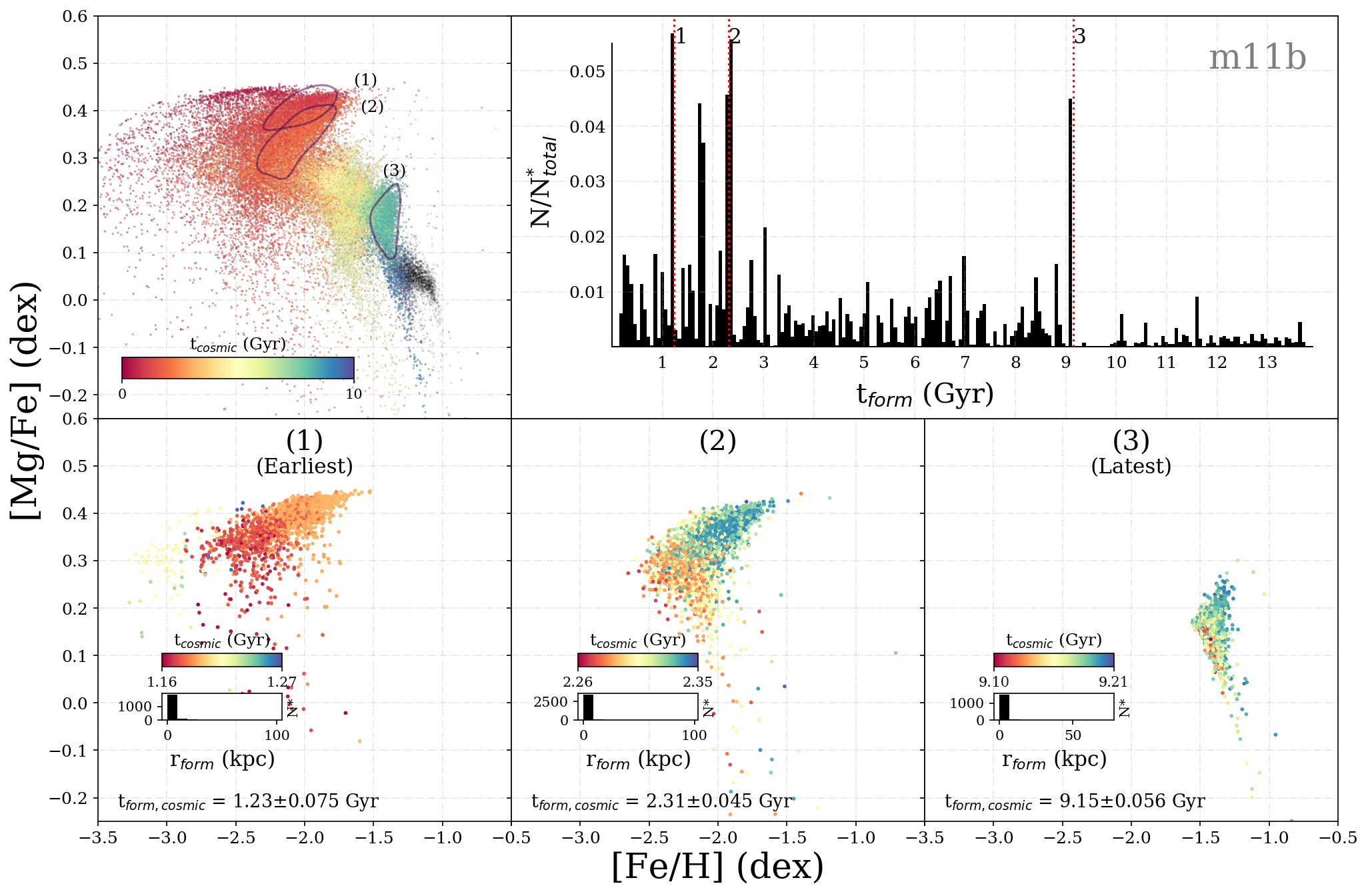}
{\label{fig:m11b_stripes}
Key aspects of \textbf{m11b}, a case study for how bursty star formation generates diagonal `striping' in elemental abundances.
\textit{Top left}: [Mg/Fe] versus [Fe/H] color-coded by time of formation (red/blue for early/late forming stars, respectively).
Numbers (1) - (3) indicate regions in [Mg/Fe] versus [Fe/H], outlined by a contour.
\textit{Top Right}: A histogram of the cosmic time at which stars formed in \textbf{m11b}.
We highlight 3 periods of star formation (red dashed lines).
\textit{Bottom Row}: We select these 3 periods and show the stars that formed within them (in the labeled ranges of cosmic time) in $\alpha$-abundance space, colored by the time of formation ($t_{\text{cosmic}}$) of stars.
All of these stars formed inside of the galaxy during individual bursts of star formation.
}

Recall that approximately $M_{*} = 5.1 \times 10^{5} \text{M}_{\odot}$ formed during the burst in Figure~\ref{fig:m10q_master}, increasing the overall stellar mass by a factor of $\sim 1.3$.
Feedback from these newly formed stars heated and blew out gas from the main galaxy, preventing further star formation for $\sim$0.66 Gyr. 
The FIRE-2 model treats a newly formed star particle as a single-age stellar population that inherits its abundances from its progenitor gas element.
As a newly formed star particle ages, its stellar population experiences core-collapse supernovae (between 3 - 37 Myr) and then Ia supernovae.
Thus, Ia supernovae significantly contributed to gas enrichment during this shut-down period.
This polluted the surrounding medium with $\alpha$-elements and some Fe.
In FIRE-2, core-collapse supernovae produce $\sim 10^{-1}$ $M_{\odot}$ of Mg and $\sim 0.7 \times 10^{-1}$ $M_{\odot}$ Fe per event, while Ia supernovae produce, $\sim 10^{-2}$ $M_{\odot}$ of Mg and $\sim 1$ $M_{\odot}$ Fe.
Thus, a Ia supernova produces nearly 100 times more Fe than Mg, so it enriches the surrounding medium with Fe.
As a result, stars that formed after the break inherit this higher fraction of Fe from their progenitor gas elements.
This manifests as an overall drop in [Mg/Fe] and an increase in [Fe/H], as the red points of Figure~\ref{fig:m10q_master} (bottom right) shows.
Stars that formed after the burst then fall in a distinct region of elemental abundances.
While distinct, this region does represent an extension of the overall trend (from upper left to bottom right).

In summary, we investigate a break in the elemental abundance trend ([Mg/Fe] versus [Fe/H]) of \textbf{m10q} (Figure~\ref{fig:m10q_master}).
Most stars in \textbf{m10q} formed before this break, with a large amount ($5.1 \times 10^{5} M_{\odot}$) forming together during a burst of star formation at $t_{\text{cosmic}} \approx 2.8 ~\text{Gyr}$.
The feedback (CCSN, SNe Ia, stellar winds) from this co-forming population significantly heats gas and pauses star formation for 0.66 Gyr, during which time SNe Ia continue to enrich the gas around the galaxy, so when star formation resumes, the increase in Fe from SNe Ia results in a drastic ($\sim 0.3$ dex) decrease in the [Mg/Fe] of newly formed stars, offset to higher [Fe/H] (as region II in Fig.~\ref{fig:m10q_master} shows).
Thus, we conclude that a strong burst in star formation can manifest as a `break' in the overall abundance trend.

\subsection{Striping of Abundance Patterns in m11b from Bursty Star Formation}
\label{m11b}

\pagefigurenv{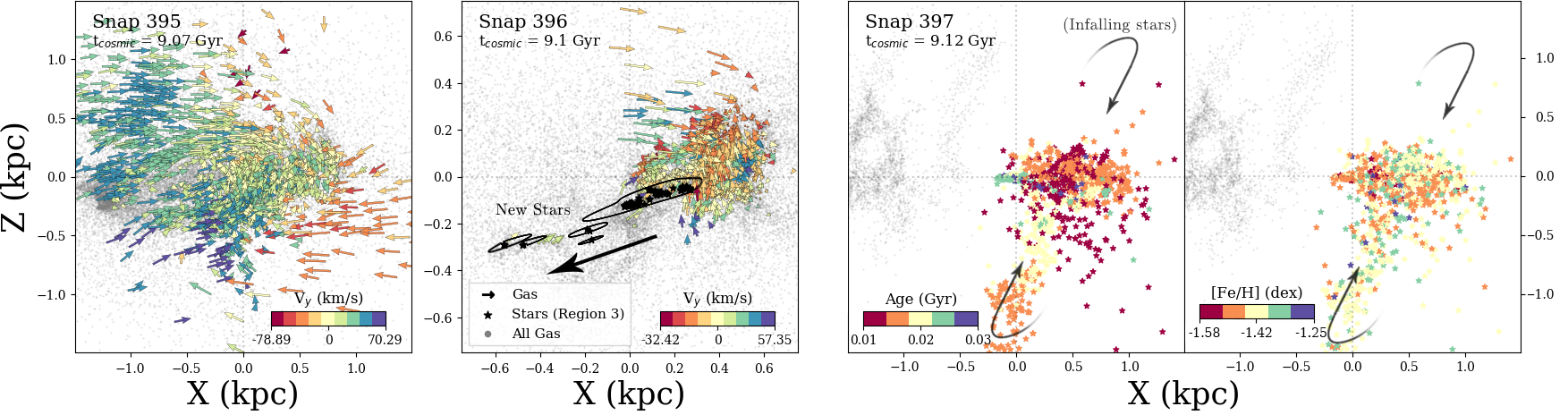}{\label{fig:m11b_star_age}
Images of gas and stars throughout a burst in star formation in \textbf{m11b}.
These images span $t_{\text{cosmic}} = 9.07 - 9.12$ Gyr (3 snapshots across $\sim 50$ Myr).
Faint gray points shows all gas elements, while arrows depict the sub-set of gas that eventually turns into stars in panel (3) of Figure~\ref{fig:m11b_stripes}.
Star-shaped points represent star particles from panel (3). 
In the left panel, we find no stars from panel (3) and the associated gas is falling inward.
In the second panel, the first stars (black star points) formed near the center. 
Feedback blows gas outward; stars that form from this gas inherit the velocity and continue outward.
In the third panel, we color all stars by their age at $t_{\text{cosmic}} = 9.12$ Gyr, as they fall back in from the burst.
Black arrows indicate the path that these star particles took; no gas from the first panel remains.
We find the oldest (dark red) near the center at present day.
These stars had additional time to fall in, while younger stars are still on infall.
The right panel shows the same stars colored by [Fe/H].
These stars formed in the main galaxy and experienced enrichment from core-collapse supernovae during the starburst.
}

Figure~\ref{fig:global_scatter} (second panel from left, bottom) shows [Mg/Fe] versus [Fe/H] for \textbf{m11b}.
This galaxy contains stripes in its elemental abundance track that run roughly perpendicular to the main trend.
In a similar manner to \textbf{m11h} and \textbf{m10q}, we color-code the time of formation of stars in the top left panel in Figure~\ref{fig:m11b_stripes}.
The stars in each stripe formed within 90 Myr of one another.
We number 3 of these stripes, corresponding to the numbered peaks in the top right panel of Figure~\ref{fig:m11b_stripes}.
There, we show the cosmic time of formation for stars in \textbf{m11b} (black bars).
We select 3 bins that correspond to strong peaks in the star formation of \textbf{m11b} (labelled 1 through 3).
While we consider 4 additional peaks, we find the same results; for brevity we only focus on 3.
Figure~\ref{fig:m11b_stripes} (bottom) shows stars born within 20-90 Myr of these peaks.
The bursts of star formation correspond with these stripes in the elemental abundances of \textbf{m11b}.

Within each individual stripes, the oldest stars are at lower values of [Fe/H] and [Mg/Fe], while younger stars at higher [Fe/H] and [Mg/Fe].
The embedded histogram(s) show the radius of formation for stars within each panel: these stars all formed \textit{in-situ} ($< 10$ kpc from the host).
As Table~\ref{table:1} shows, nearly a quarter of stars in \textbf{m11b} formed ex-situ, but the stars in these stripes all formed inside the main galaxy during single bursts of star formation and rapid enrichment.

Figure~\ref{fig:m11b_star_age} shows the positions of gas elements that turn into the stars particles in panel (3) of Figure~\ref{fig:m11b_stripes} across 3 timesteps ($t_{\text{cosmic}} = 9.07, 9.1, 9.12 \, \text{Gyr}$).
Arrows show the direction of motion of gas, and their lengths correspond to the magnitude of their velocity in the x-z plane ($v_{\text{xz}}$), while their color indicates $v_{\text{y}}$.
The faint gray points show the total gas distribution, and stars show star particles.

In Figure~\ref{fig:m11b_star_age} (left panel), we find no stars, while gas is falling towards the center.
In the second panel, the gas continues to collapse and most is within $\sim 0.7$ kpc.
The first stars (encircled in black) from this sub-selection formed near the center, and are all moving outwards from feedback that pushed their parent gas particles (as indicated by the black arrow).

Finally, the third and fourth panels show that all of the gas particles corresponding to panel (3) have turned into stars, which are now falling towards the galactic center (as black arrows indicate).
The third panel shows the stars colored by their age at 9.12 Gyr.
The oldest stars (green, blue) are closer to the center than some of the younger stars (red, orange), which are found as far out as 1 kpc.
The fourth panel shows these same stars colored by their [Fe/H].
The oldest stars in the third panel correspond to the most metal-poor stars in the fourth panel.
These stars formed within 50 Myr of one another, during which time CCSN dominated the feedback and enrichment (across $\sim 3 - 37$ Myr), leading to rapid Mg enrichment.

This process is identical for the bursts in panels (1) and (2) of Figure~\ref{fig:m11b_stripes}; after all of the stars in a given stripe have formed, the youngest stars from the burst are at larger radii than the oldest, which have had more time to fall into the center.
This entire process is a recurring cycle throughout \textbf{m11b}'s history; bursts of star formation drove gas out, which then cooled and fell back into the main galaxy, forming stars near the center (as described in \citet{Alcazar2017}).
Additionally, \citet{El-Badry2016} highlighted this process for several low-mass galaxies in FIRE, and we find that the signature of these bursts is a sharp increase in [Mg/Fe] and smaller increase in [Fe/H].
This differs from the effects of the burst in \textbf{m10q}, in which its lower total mass combined with an exceptionally strong burst led to a pause in star formation and a break in the abundance distribution.
Given the snapshot time resolution of $\sim 20$ Myr, we are not able to probe the geometry of how these stars formed in detail.
\citet{Yu2020} performed a detailed investigation of stars forming in the outflows of MW-mass galaxies in FIRE-2; cool, high-density regions form as a result of supernova feedback and become self-gravitating, leading to star formation.
We discuss the importance of this to observational findings in \S\ref{sec:conc}.
However, it remains unclear whether the stars in \textbf{m11b} formed alongside the gas propagating out from the center, or together in the galaxy before being blown out.
We will investigate this in further detail in future work.

Overall, we find stripes orthogonal to the main elemental enrichment trend in the [Mg/Fe] versus [Fe/H] in \textbf{m11b} (Figure~\ref{fig:m11b_stripes}).
Stars across individual stripes follow a positive trend in [Mg/Fe] versus [Fe/H]; they all formed within 100 Myr of each other during the same burst, during this time CCSN rapidly enriched the gas with Mg.
These stars formed \textit{in-situ} -- within 10 kpc of the main galaxy -- and were not accreted.
Bursty star formation combined with its feedback drove gas out from the center of the galaxy, but it eventually fell back in, driving further star formation.
Thus, these stripes are signatures of recurrent bursts in star formation in \textbf{m11b}.

\pagefigurenv{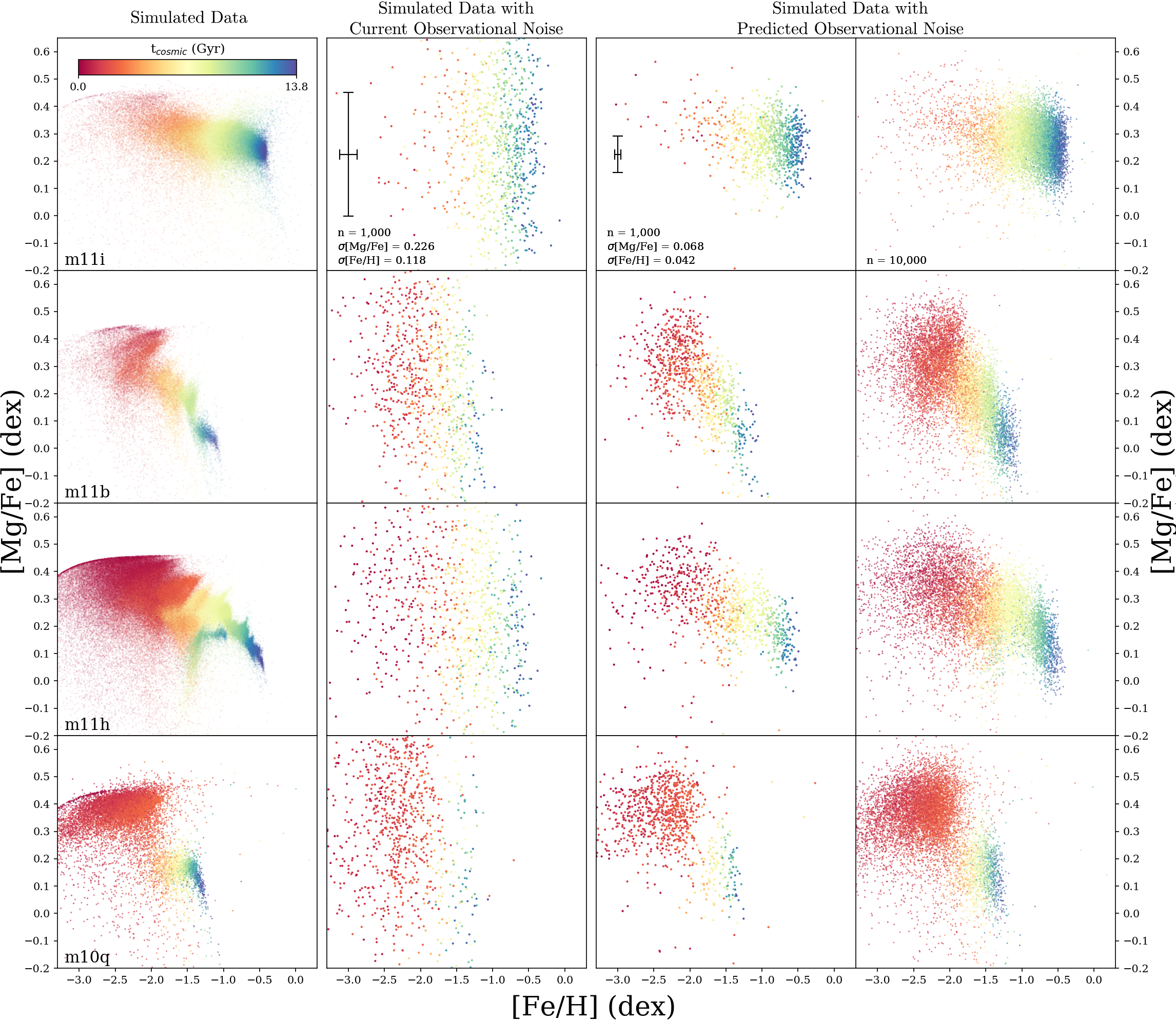}{\label{fig:obs}
[Mg/Fe] versus [Fe/H] for galaxies that we explored as case studies: \textbf{m11i} (top row), \textbf{m11b}, \textbf{m11h}, and \textbf{m10q} (bottom row).
\textit{Left}: points show star particles in the galaxy at $z = 0$, colored by the cosmic time at which they formed.
We see little complexity in \textbf{m11i};
\textbf{m11b} shows stripes that run orthogonal to the main track in; \textbf{m11h} shows a secondary track; and \textbf{m10q} shows a break in its track.
\textit{Second column}: Synthetic measurements that mimic current observations; we randomly sample 1,000 star particles, and we add scatter to each point by randomly drawing from a Gaussian whose $1$-$\sigma$ uncertainty in [Mg/Fe] and [Fe/H] correspond to the average uncertainties of those abundances from the data in Table 3 for Ursa Minor in \citet{Kirby_2018}.
\textit{Right two columns}: Synthetic future observations; we randomly sample 1,000 and 10,000 star particles, and we add in additional scatter based on predicted observational uncertainties for James Webb Space Telescope (JWST) from Figure~12 in \citet{Sandford2020}, which are within 0.05 dex of the predicted observational uncertainties for the Extremely Large Telescope(s) (ELTs).
We enlarge the point size in the middle two columns by a factor of 10 for visibility.
}

\section{Observability} 
\label{sec:Observability}

\subsection{Synthetic Current Observations}
\label{current}

Given our results above, an important question is: how measurable are these predicted features in elemental abundances if one includes typical observational uncertainties?

Figure~\ref{fig:obs} (left) shows [Mg/Fe] versus [Fe/H] for star particles in \textbf{m11i} (top), \textbf{m11b}, \textbf{m11h}, and \textbf{m10q} (bottom) at present day, as we explored above.
We generate mock observations of these stars by adding observational uncertainties in measurements of [Mg/Fe] and [Fe/H] from \citet{Kirby_2018}, who catalog several elemental abundances for over a thousand stars in nearby dwarf galaxies.
Specifically, we incorporate the quoted uncertainties from Ursa Minor dwarf galaxy, because it is relatively close to the Milky Way ($\sim 60$ kpc), and it is within the mass range of our simulated sample.
This galaxy is well sampled in measurements of [Mg/Fe] and [Fe/H] (146 stars).
We generate mock uncertainties by sampling values from a Gaussian probability distribution of the form:
\begin{equation}
    p(x) = \frac{1}{\sqrt{2 \pi \sigma^2}}\text{exp}{\left( -\frac{x^2}{2\sigma^2} \right)},
\end{equation}
with $\sigma_{\text{[Fe/H]}}$ = 0.11 dex or $\sigma_{\text{[Mg/Fe]}}$ = 0.22 dex, which we take from Table 3 in \citet{Kirby_2018}.
We then associate these sampled values with randomly selected star particles from our simulations, and we couple them as errors on the original simulated values of [Mg/Fe] and [Fe/H].

We randomly sample uncertainties from the distributions above and incorporate them as uncertainties in the values of [Mg/Fe] and [Fe/H] for the second column in Figure~\ref{fig:obs}.
We choose $\text{N = 1,000}$, because it is an ambitious but feasible number of stars for current/near-term observations, given  measurements of stars in nearby dwarf galaxies today.
For example, \citet{Kirby_2018} provide spectroscopic measurements of abundances for 562 stars in Leo I and 403 stars in Fornax.
We sample a Gaussian with a standard deviation ($\sigma$) equal to the average measurement error $\text{N}$ times. 
From this, we obtain a random subset of our data with N points and the same scatter as observations (Figure~\ref{fig:obs}, 2nd column). 
The color gradient representing the $t_{\text{form}}$ of these stars still progresses from older stars at a low [Fe/H] to young stars at a high [Fe/H] in all galaxies.
However, none of the features that we detailed in \S\ref{sec:res} are distinguishable under current observational uncertainties.
We also generate this distribution under current observational constraints for 10,000 stars (not shown), finding similar result, meaning that it is the measurement uncertainties, and not the number of stars sampled, that obfuscates these features.

Overall, we still are able to identify the primary elemental abundance trends (from top-left to bottom-right), but the secondary structures are obscured.
Therefore, current observations with their uncertainties can provide general constraints on a galaxy's history based on these trends, but even 10,000 data points from a single galaxy are insufficient to detect the types of complex features that we analyzed in \S\ref{sec:res}.
We conclude that decisively detecting such features in nearby galaxies and leveraging them to understand galactic history is challenging with current observational uncertainties.



\subsection{Synthetic Future Observations}
\label{jwst}

Upcoming spectroscopic surveys like James Webb Space Telescope (JWST, \citet{Gardner2006}), Prime Focus Spectrograph on Subaru (PFS, \citet{Sugai2012}), 4-metre Multi-Object Spectroscopic Telescope (4-MOST, \citet{Daan2011}), Thirty-Metre Telescope (TMT, \citet{Nelson2008}), Giant Magellan Telescope (GMT, \citet{Bernstein2018}), and the European-Extremely Large Telescope (E-ELT, \citet{deZeeuw2014}) will provide higher precision measurements of abundances (discussed further in \S\ref{sec:conc}).
For the purposes of this analysis, we use predictions of uncertainties for JWST and the E-ELTs. 
Using the Cramer-Rao Lower Bound (CRLB) estimate,\footnote{CRLB is a statistical method that allows one to estimate the minimum theoretical variance of a quantity; it is useful for feasibility studies. In this paper, the variances of interest are for [Mg/Fe] and [Fe/H].} \citet{Sandford2020} provide predictions for uncertainties on abundances in future observations like JWST and the Extremely Large Telescopes.
We apply these minimum uncertainties to our simulation data using a similar method to \S\ref{current}; we determine whether upcoming surveys could measure our predicted features in elemental abundances.
Figure~\ref{fig:obs} (right 2 columns) shows [Mg/Fe] versus [Fe/H] for 1,000 and 10,000 randomly selected star particles.
We choose 10,000 for the final column, because it is an optimistic projection for the number of stars that one could measure over the upcoming decade.
Using the same methodology from \S\ref{current}, we add randomly drawn uncertainties generated through a Gaussian distribution whose $1-\sigma$ corresponds to the average of uncertainties proposed for measurements on abundances for [Mg/Fe] and [Fe/H] in Figure 12 of \citet{Sandford2020} ($\sigma_{\text{[Fe/H]}}$ = 0.0425 dex and $\sigma_{\text{[Mg/Fe]}}$ = 0.0675 dex).

These synthetic observations resolve the general trends in [Mg/Fe] versus [Fe/H] and age.
Unlike the second column, we find a much tighter distribution in [Mg/Fe] for a given [Fe/H] around all of the tracks (that is, there is a $\sim 0.3$ dex spread in [Mg/Fe] for \textbf{m11i} in the third column versus a $\sim 0.7$ dex spread in the second column). 
In the case of \textbf{m11i}, the cutoff at [Fe/H] $= -1.0$ is visually much sharper than it was in the second column.
In \textbf{m11b}, the characteristic sharp rises in [Mg/Fe] (stripes) are still not distinguishable.
In \textbf{m11h}, the secondary track is not visible, though a small number of the young stars from the secondary track are visible at lower metallicities ([Mg/Fe] $\sim$ 1.0).
In \textbf{m10q}, the break is visible at [Fe/H] $\sim$ -2.0.
When looking at the right column (N = 10,000), the break in \textbf{m10q} becomes even more visually distinct.
Under these uncertainties, this `break' is the only visible feature.

With the precision of upcoming spectroscopic observations, like JWST and the E-ELTs, we may detect features that signify temporary quenching events (like the break in \textbf{m10q}) in low-mass galaxies.
Measuring such features would provide critical insight to the formation histories of these galaxies, complementary to star-formation histories derived from modeling the color-magnitude diagram of resolved stellar populations.
However, it remains unclear if such observations could resolve the features we explored from \textbf{m11h} or \textbf{m11b}.

\section{Summary and Discussion}
\label{sec:conc}
\subsection{Summary}

We examined trends in [Mg/Fe] versus [Fe/H] for stars in 8 FIRE-2 cosmological simulations of low-mass galaxies (Figure~\ref{fig:global_scatter}).
In all galaxies, we find that the overall trend in elemental abundances goes from low [Fe/H] and high [Mg/Fe] to high [Fe/H] and low [Mg/Fe],
which persists self-similarly at all radii.
However, this trend is not smooth and monotonic in half (4 of 8) of the galaxies.
\textbf{m11h} and \textbf{m10q} contain multiple enrichment tracks sourced from satellite accretion and bursty star formation.
Furthermore, \textbf{m11b}, \textbf{m11h}, and \textbf{m11q} contain `stripes' that run orthogonal to the main trend, indicating recurrent bursty star formation.
We investigate 3 (\textbf{m11h, m10q, m11b}) of these galaxies in detail:

\begin{itemize}
  \item \textbf{m11h} shows a secondary pattern in [Mg/Fe] versus [Fe/H] (Figure~\ref{fig:m11h_master}). 
  This feature was the result of a 1:3 mass-ratio galaxy merger (Figure~\ref{fig:m11h_3column}) that was gas-poor and accreted stars onto the main galaxy.
  The feature persists at all radii, though it is most distinct 2 kpc to 4 kpc from the center (Figure~\ref{fig:m11h_radial}).
  Other galaxies (\textbf{m10q}, \textbf{m11e}) contain similar features, and we investigate \textbf{m10q} in detail (\S\ref{m10q}).
  
  \item \textbf{m10q} also shows a break (bimodality) in its [Mg/Fe] versus [Fe/H] (Figure~\ref{fig:m10q_master}).
  Unlike \textbf{m11h}, this break does not arise from a merger, but from a large burst in star formation that blew out gas and halted star formation for $\sim 0.66$ Gyr.
  During this time, enrichment from Ia supernovae dominated; when star formation resumed, new stars formed at lower [Mg/Fe] and higher [Fe/H].

  \item \textbf{m11b} shows several stripes in [Mg/Fe] versus [Fe/H] that run roughly orthogonal (increasing [Mg/Fe] with increasing [Fe/H]) to the overall trend (Figure~\ref{fig:m11b_stripes}).
  The stars in a given stripe formed \textit{in-situ}, around the same time, and during a single burst in star formation.
  The recurrence of these stripes arose from recurrent bursty star formation in \textbf{m11b}; during a given burst, feedback drove out gas while star formation enriched the surrounding medium.
  This led to a characteristically sharp rise in the [Mg/Fe] versus [Fe/H] that appears as a `stripe'.
  This striping is visible in the elemental abundance tracks of \textbf{m11h} and \textbf{m11q} as well, but we used \textbf{m11b} as our case study, because its stripes are most distinct.
\end{itemize}

Thus, we found breaks in the elemental abundance trends for 2 (of 8; 25\%) galaxies, which were either a separate elemental enrichment track from a galaxy merge or a split in the overall trend as a result of a strong burst in star formation.
Stripes that run perpendicular to the main trend were clear in 3 (of 8; 37.5\%) of our galaxies.
In total, 50\% (4 out of 8) of galaxies in our suite contained features in their elemental abundance patterns that are the elemental imprints of prior mergers or bursty star formation.s

We then addressed the feasibility of observing such features using spectroscopic surveys:

\begin{itemize}
    \item We considered current observational uncertainties for [Mg/Fe] and [Fe/H] in Figure~\ref{fig:obs} taken from spectroscopic observations of stars in Ursa Minor.
    We randomly selected 1,000 star particles from the simulation and applied these uncertainties, finding that these uncertainties are too large to discern any of the notable features we explored.

    \item Similarly, in the right two columns of Figure~\ref{fig:obs} we considered projected observational uncertainties for JWST.
    Here, the original abundance features are easier to resolve.
    For 1,000 and 10,000 observations per galaxy, the break in \textbf{m10q} is visually distinct.
    Furthermore, for \textbf{m11h} we see the younger population of stars in the secondary track, though the gap in [Mg/Fe] is not distinguishable.
\end{itemize}

\subsection{Discussion}

Other works have used simulations to investigate the connection between critical events in a low-mass galaxy's formation history and the abundances of its stellar populations.
\citet{Zolotov2010} used \textsc{Gasoline} simulations to map the abundances of accreted stars to a distinct population in $\alpha$-versus-Fe (similar to our analysis for \textbf{m11h}).
\citet{Genina2019} used the \textsc{APOSTLE} simulations to understand the tendency for higher-metallicity stars to populate the center of the galaxy while lower-metallicity stars extend to larger radii \citep[for example][]{Tolstoy2004}.
They explored how this spatial segregation forms, finding that accretion and feedback processes are integral to the formation of the feature.
\citet{Buck2021} examined \textsc{Gasoline2} simulations and reported a diversity in the $\alpha$-abundance patterns of their galaxies, some with comparable stellar masses to those in our study.
They comment that feedback processes (nucleosynthetic yields, Ia time delays) and accretion play a major role in formulating this diversity.
We expand on this type of analysis and find that evidence of bursty star formation is observable in elemental abundances of stars as well (\S\ref{sec:res}).

Previous works have examined the influence of bursty star formation in low-mass galaxies in FIRE-2.
For example, \citet{Chan2015, Chan2018} detailed the dynamical effects of such feedback in FIRE-2 for halos with masses $10^{10} - 10^{11} M_{\odot}$ (including \textbf{m11b}).
Star particles in FIRE-2 inherit velocities from progenitor gas elements during outflows, and they find that such a process helps reproduce properties (enclosed mass, mass-to-light ratio) consistent with observed ultra-diffuse galaxies.
\citet{El-Badry2016} identified `breathing modes' driven by bursty star formation that push stars to dispersion-dominated orbits over time.
\citet{Sparre2017} investigated the relationship between simulated star-formation rates (SFR) and stellar mass (M$_*$), noting that the burstiness of star formation in FIRE affects the scatter in SFR-M$_*$.
\citet{Alcazar2017} showed how bursty stellar feedback drives outflows that recycle multiple times, and which can help sustain star formation in dwarf galaxies at late times.
Overall, bursty star formation plays a major role in the development low-mass galaxies through stellar feedback.

In general, accretion and feedback processes lead to distinct stellar populations in the [Mg/Fe] versus [Fe/H] of simulated galaxies in FIRE, \textsc{Apostle}, and \textsc{Gasoline2} simulations.
Features like the striping in \textbf{m11b} and break in the trend of \textbf{m10q} are the result of feedback processes as well, while the secondary track in \textbf{m11h} is from a galaxy merger. 
The prevalence of these features in present-day elemental abundance distributions substantiates the importance of accretion and feedback in low-mass galaxy formation.

Observationally, \citet{Silk1987} provided an early connection between the enrichment of stars in dwarf galaxies (irregulars and ellipticals) and their evolutionary history.
They focus on high-redshift galaxies and the effects of enrichment on morphology, and also describe processes that qualitatively match our analyses.
For example, they describe how the enrichment of gas from stellar feedback can influence later stellar abundances as the enriched gas falls in and forms new stars.
This generally agrees with our analysis of \textbf{m11b}, where bursty formation leads to early feedback that both enriches and blows out gas, which later falls in to form stars at higher $\alpha$-abundances. 
\citet{Koch2006} found signs of episodic star formation in Carina dwarf galaxy.
Stars formed during bursts followed by a periods of little to no star formation.
They described a possible mechanism for this which relates to our description of \textbf{m11b}; the repeated onset and pause in star formation could be due to the re-accretion of previously blown out (but not blown away) gas. In addition, \citet{Silk1987} described how feedback from newly formed stars could disrupt star formation as a whole in these systems, leading to a pause in star formation.
This agrees with \citet{Koch2006}, in that star formation would not resume if the gas was blown away.
Furthermore, it compliments our analysis for \textbf{m10q}, where we focus on a single burst in star formation leading to feedback that is disruptive enough to pause star formation all together for 0.66 Gyr.

Observations suggest that star formation can be associated with outflowing gas.
For example, \citet{Maiolino2017} directly observed star forming outflows in an external galaxy (IRAS F23128-5919).
\citet{Gallagher2019} repeated their analysis with a large sample of observed galactic outflows (N = 37), and found that atleast $\sim30 \%$ of the outflows are actively forming stars.
Galaxies in FIRE-2 are consistent with these findings; stars form within dense outflows in several simulations \citep{Yu2020}.
This offers some insight into the geometry of star formation during a burst in \textbf{m11b}, as it indicates that stars can form in the outflows following a burst.

\citet{Koch2008} found a wide range in the [$\alpha$/Fe] for some stars in Carina and postulated that it could be the imprint of an early accretion event.
Similarly, in \textbf{m11h}, we see a population of stars at distinctly lower [Mg/Fe] for a given value of [Fe/H] (we refer to this as a secondary track, or a `break' in the overall trend).
While they cite different magnitudes for the gap in $\alpha$-abundance than we do ($\sim 0.5$ dex versus $\sim 0.1-0.2$ dex), qualitatively we agree: a gap in the [$\alpha$/Fe] versus [Fe/H] could indicate accreted material because of a separate star formation history. 
\citet{Ruiz-Lara2020} find 2 metal-poor populations of stars in Leo I that are thought to be accreted material, indicated by their differing metal-content (as well as the SFH of Leo I).

Other observational works identify $\alpha$-abundances and metallicities in nearby ultra-faint dwarf galaxies (Carina II and III, Grus I, Triangulum II), including trends in $\alpha$-verus-iron and age \citep[see:][]{Ji2018, Ji2019}.
While the galaxies in our study are more massive than ultra-faints, ongoing studies of incrementally fainter sources paint an optimistic picture for sharply resolving the spectra of more stars in nearby dwarf galaxies.
Currently, surveys like Stellar Abundances for Galactic Archaeology (SAGA) and observations from Keck present the abundances for thousands of stars in nearby dwarf galaxies \citep[e.g.][]{Suda2017, Kirby_2018}. 
In the near future, instruments like the Subaru Prime Focus Spectrograph (PFS), the 4-metre Multi-Object Spectroscopic Telescope (4-MOST), and JWST/ELTs will gather stellar spectra in nearby galaxies \citep{Daan2011, Takada2014}.
This will enable further analysis of trends in elemental abundances.

Several works also investigate elemental abundance distributions of observed galaxies and evaluate their histories. 
\citet{Webster2014} find that large scatters in the [$\alpha$/Fe] distributions for nearby dwarf galaxies (i.e. Ursa Major I) are best explained by bursty histories, as opposed to a single episode of star formation.
\citet{Vargas2014, Suda2017} look at low-mass systems and focus on the identification of the `$\alpha$-knee', or the metallicity at which there is a turnover in $\alpha$-abundance driven by the onset of SNe Ia.
The metallicity of this turnover is linked to early star formation rates in galaxies.
\citet{Nidever2020} identify a rise in the [Mg/Fe] versus [Fe/H] of stars in the LMC, and associate it with a recent burst in star formation as well.
These works show that analyzing the elemental abundances is increasingly feasible for unraveling the histories nearby dwarf galaxies, as we point to similar rises in the [Mg/Fe] of stars in \textbf{m11b} to explain its history. 

Ultimately, observational uncertainties and sample size determine whether we can find these features in observed low-mass galaxies.
We present mock observed abundances for 1,000 and 10,000 stars in \S\ref{sec:Observability} to provide some insight into these constraints.
Present-day observational uncertainties from \citet{Kirby_2018} do not permit the recovery of any of the features we detail in this paper, despite a relatively large sample of over 100 stars.
However, \citet{Bedell2014} assert that even with the instruments that are used in the current era, sub-0.01 dex precision could be achieved for some sources with careful selection criteria; while this would reduce the sample size significantly, it speaks to the potential for higher-resolution in abundance measurements.
\citet{Sandford2020} provide predictions for uncertainties of such measurements in JWST/ELTs.
Using their analysis, we derive expected future uncertainties for [Mg/Fe] and [Fe/H] (0.068 and 0.042).
This uncertainty is still larger than the potential precision provided by \citet{Bedell2014}, but it generally agrees with an expectation for higher precision and sample size with time, permitting us to observe such features. 
Furthermore, we find that these uncertainties are small enough to resolve the break in \textbf{m10q}.

Currently, sample sizes and measurement uncertainties of elemental abundances restrict us from observing the imprints of accretion and bursty star formation in [Mg/Fe] versus [Fe/H].
Studies like APOGEE contribute significantly to the volume and precision of current stellar spectroscopic data, enabling notable work on stellar abundances in nearby galaxies like the LMC and Ursa Minor.
Upcoming studies like JWST, PFS, and 4-MOST will continue to increase sample size and refine precision in measurements of stars in nearby galaxies.
Thus, this analysis may become a viable means for investigating the formation history of low-mass galaxies.
While our analysis indicates that such features are not uncommon in the FIRE-2 model, our current sample size is limited to 8 simulated galaxies and a larger suite would provide stronger statistical predictions.
Nonetheless, our results argue that mergers/accretion and bursty star formation have distinguishable impacts on the present-day abundance distributions of even low-mass galaxies, so measurements of these features can inform us about such galaxies' formation histories.


\section*{Software Used}
\label{sec:Software}
We used the publicly available Python packages, \textsc{GizmoAnalysis} \citep{Wetzel_Gizmo} and
\textsc{HaloAnalysis} \citep{Wetzel_Halo} to analyze these data.

\section*{Data Availability}
\label{sec:Data}
All Python code that we used to generate these figures, including the data content of the figures, is available at \url{https://github.com/patelpb96}.
FIRE-2 simulations are publicly available \citep{Wetzel2022} at \url{http://flathub.flatironinstitute.org/fire}.
Additional FIRE simulation data is available at \url{https://fire.northwestern.edu/data}.
A public version of the \textsc{Gizmo} code is available at \url{http://www.tapir.caltech.edu/~phopkins/Site/GIZMO.html}.

\section*{Acknowledgments}
\label{sec:acknowledgements}

We thank the anonymous referee, Jenna Samuel, and Anna Parul for useful comments.
Support for PBP was provided by the the Blue Waters sustained-petascale computing project, which is supported by the National Science Foundation (awards OCI-$0725070$ and ACI-$1238993$) and the state of Illinois. Blue Waters is a joint effort of the University of Illinois at Urbana-Champaign and its National Center for Supercomputing Applications.
AW received support from:: NSF via CAREER award AST-2045928 and grant AST-2107772; NASA ATP grants 80NSSC18K1097 and 80NSSC20K0513; HST grants GO-14734, AR-15057, AR-15809, GO-15902, GO-16273 from STScI; a Scialog Award from the Heising-Simons Foundation; and a Hellman Fellowship.
CAFG was supported by NSF through grants AST-1715216, AST-2108230,  and CAREER award AST-1652522; by NASA through grant 17-ATP17-0067; by STScI through grant HST-AR-16124.001-A; and by the Research Corporation for Science Advancement through a Cottrell Scholar Award.
We performed this work in part at the Aspen Center for Physics, supported by NSF grant PHY-1607611.
We ran simulations using: XSEDE, supported by NSF grant ACI-1548562; Blue Waters, supported by the NSF; Pleiades, via the NASA HEC program through the NAS Division at Ames Research Center.




\bibliographystyle{mnras}
\bibliography{bibliography} 




\bsp	
\label{lastpage}
\end{document}